\documentclass[11pt]{article}
\usepackage[top=1in,bottom=1in,left=0.75in,right=0.75in]{geometry}
\usepackage{graphicx}  
\usepackage{xcolor} 
\usepackage{amsmath,amssymb}
\usepackage{tabularx}
\usepackage{authblk}
\usepackage{newtxtext,newtxmath}

\makeatletter
\renewcommand{\maketitle}{%
  \begin{center}
    \vspace*{-1.5em} 
    {\large\bfseries \@title \par} 
    \vspace{0.5em}
    {\normalsize \@author \par} 
  \end{center}
  \vspace{1em} 
}
\makeatother

\begin{document}

\title{\bfseries Anisotropic fully-gapped superconductivity in quasi-one-dimensional Li$_{0.9}$Mo$_6$O$_{17}$}

\date{} 

\author[1]{M. J. Grant}
\author[1]{T. M. Huijbregts}
\author[1]{R. Nicholls}
\author[2]{M. Greenblatt}
\author[3,4]{P. Chudzinski}
\author[1]{A. Carrington}
\author[1,5]{N. E. Hussey}

\affil[1]{H. H. Wills Physics Laboratory, University of Bristol, Bristol, BS8 1TL, UK}
\affil[2]{Department of Chemistry and Chemical Biology, Rutgers University, Piscataway NJ, USA}
\affil[3]{School of Mathematics and Physics, Queen's University Belfast, Belfast, UK}
\affil[4]{Institute of Fundamental Technological Research, Polish Academy of Sciences, Warsaw, Poland}
\affil[5]{High Field Magnet Laboratory (HFML-FELIX) and Institute for Molecules and Materials, Radboud University, Nijmegen, Netherlands}

\maketitle
\begin{abstract}
Superconductivity in quasi-one-dimensional Li$_{0.9}$Mo$_6$O$_{17}$ emerges from an exotic, non-metallic normal state that exhibits signatures of Tomonaga-Luttinger liquid behavior, emergent symmetry and excitonic order. The high upper critical field, $H_{c2}$, in Li$_{0.9}$Mo$_6$O$_{17}$ suggests that that the favored pairing state is spin-triplet in nature. Here, we report measurements of the magnetic penetration depth down to $0.08\,\mathrm{K}$ ($T/T_c \lesssim 0.04$) and the specific heat down to $0.4\,\mathrm{K}$ ($T/T_c \lesssim 0.2$), and show that they are consistent with a moderately-coupled, fully-gapped superconducting state with marked gap anisotropy and a minimum ($\Delta_{\rm min} \simeq 0.4\,k_{\mathrm{B}}T_c$) occurring over a very narrow region in $k$-space. Combined with knowledge of $H_{c2}$, these measurements support the presence of a nodeless and possibly odd-parity spin-triplet superconducting order parameter in Li$_{0.9}$Mo$_6$O$_{17}$. 

\end{abstract}

\textit{Introduction} - Electrons confined to propagate along weakly coupled chains are highly susceptible to different forms of electronic order \cite{dressel2007}. Such quasi-one-dimensional (quasi-1D) conductors offer a promising platform for the realization of exotic phases, such as the Tomonaga-Luttinger liquid (TLL) state \cite{giamarchi2004} and spin-triplet superconductivity \cite{zhang2007}. Materials that exhibit both superconductivity and TLL behavior are rare, though carbon nanotubes and the Bechgaard salt (TMTSF)$_2$PF$_6$ are potential candidates. In the former, intrinsic superconductivity is only realized in bundles (ropes) of nanotubes \cite{kociak2001}, while in the latter, superconductivity appears under applied pressures of around 12 kbar \cite{jerome1980}. Hence, in both cases, it is not clear whether the TLL state survives as the coupling between adjacent conducting chains becomes stronger and superconductivity emerges.

Evidence for unconventional (nodal \cite{pang2015} and triplet \cite{yang2021}) superconductivity as well as TLL behavior \cite{watson2017} has also been reported in a family of quasi-1D arsenide superconductors with the generic formula A$_2$X$_3$As$_3$ (A = Na, K, Rb, Cs, and X = Cr, Mo)  \cite{bao2015}. However, the electrical resistivity anisotropy and thus the degree of one-dimensionality in this family is not yet known, and may in fact be compromised by the presence of additional 3D components in its electronic structure \cite{watson2017}. Certainly, $H_{c2}$ exhibits only modest anisotropy \cite{balakirev2015}.

Li$_{0.9}$Mo$_6$O$_{17}$ (LMO, or purple bronze) is another viable candidate for the realization of a bulk TLL that also superconducts \cite{popovic2006, chudzinski2017}. Supporting evidence for TLL behavior in LMO includes power-law scaling of the low-energy density of states as measured by scanning tunnelling spectroscopy and angle-resolved photoemission spectroscopy \cite{podlich2012, denlinger1999, hager2005, wang2006, dudy2012}, and a marked violation of the Wiedemann-Franz law, even at $300\,\mathrm{K}$ \cite{wakeham2011, cohn2012}. Its basic electronic structure -- a pair of 1D sheets with $d_{xy}$ character -- derives from the MoO$_6$ octahedra that form conductive zig-zag chains along the $b$-axis \cite{onoda1987, popovic2006}. Accordingly, the resistivity of LMO is highly anisotropic with $\rho_a$:$\rho_b$:$\rho_c \approx$ 100:1:2000 \cite{mercure2012, ke2021}. Upon cooling, $\rho_b(T)$ is metallic ($d\rho/dT>0$) down to a minimum at $T_{\mathrm{min}}\simeq 25\,\mathrm{K}$, below which there is a crossover to insulating behavior in the form of a large upturn in $\rho(T)$ before the onset of superconductivity at $T_c \sim$ 2 K. The collapse of this resistive upturn under an applied magnetic field \cite{xu09}, as well as the absence of any definitive thermodynamic or spectroscopic signatures at $T_{\mathrm{min}}$ \cite{matsuda1986, chakhalian2005, choi2004, wu2016}, appear to rule out charge- or spin-density wave order as the origin of the upturn. More recently, it has been suggested \cite{chudzinski2017, lu2019, chudzinski2023} that the filled and empty $d_{yz}$ and $d_{xz}$ bands may combine to form excitonic states that are not optically active. These \lq dark' excitons then act as strong scattering centers whose influence grows with decreasing $T$, driving the system towards a state of \lq emergent symmetry' in which the (Mott) insulating and superconducting (SC) states are effectively degenerate \cite{chudzinski2023}.

Despite the fact that LMO is an unusual, if not unique, example of a quasi-1D superconductor emerging out of an insulating state, very little is known about the nature of its superconductivity, in particular its gap structure. To address this, we have measured the low-$T$ specific heat and penetration depth in LMO crystals with the highest-reported $T_c$ values. The combined data are found to be consistent with a nodeless SC order parameter, albeit one possessing strong ($\sim 7\times$) anisotropy along the 1D Fermi sheets. The 5-fold enhancement of $H_{c2}$ beyond the Pauli paramagnetic limit in LMO \cite{mercure2012} has led to suggestions that its pairing state is spin-triplet \cite{sepper13, cho2015, lera2015, platt2016}. The anisotropic, fully-gapped state reported here shares features with some of these models \cite{sepper13, cho2015} and while no direct evidence for triplet pairing has yet been reported, the totality of the data appear consistent with a nodeless, triplet pairing state \cite{sepper13, cho2015, lera2015, platt2016}.

\begin{table}
	\small
	\centering
	\begin{tabularx}{0.45\textwidth}{X|X|X}
		\hline\hline
	{}	& \textbf{\#1}  & \textbf{\#2} \\ \hline
         {$a$ (\AA)} & {$9.481(3)$} & {$9.511(8)$} \\
        {$b$ (\AA)} & {$5.527(4)$} & {$5.520(3)$} \\
         {$c$ (\AA)} & {$12.701(8)$} & {$12.733(13)$} \\
         {$\theta$ ($^\circ$)} & {$90.57(2)$} & {$90.68(6)$} \\
        \hline\hline
	\end{tabularx}
    \caption{Lattice parameters ($a,b,c$) and monoclinic angle ($\theta$) determined by x-ray diffraction of two LMO crystals.}\label{tb:XRD}
\end{table}

\textit{Experimental} - Single crystals were grown using a temperature gradient flux method \cite{mccarroll1984}. Crystals with a range of $T_c$ values, including non-SC samples, are found within a single growth \cite{schlenker1985,boujida1988}. The origin of this variability is not presently known \cite{cohn2012_2}, though may be associated with the near-degeneracy of the Mott insulating and SC states as reported in Ref.~\cite{chudzinski2023}. For this study, we selected samples with the highest $T_c$ values ($\gtrsim 2\,\mathrm{K}$). The crystallinity of two of the samples used for magnetic penetration depth measurements, \#1 ($\sim 580\times 350\times40\,\mu\mathrm{m}^3$) and \#2 ($\sim 310\times 270\times15\,\mu\mathrm{m}^3$) was checked using a \textit{Bruker D8 Venture} four-circle x-ray diffractometer in conjunction with a \textit{Bruker CPAD} detector. The lattice parameters obtained are shown in Table \ref{tb:XRD}, and are in good agreement with previous reports \cite{onoda1987, ke2021}.

Measurements of the $T$-dependence of the magnetic penetration depth, $\Delta\lambda(T)$ were carried out on three samples (\#1, \#2, \#3) in a dilution refrigerator down to $80\,\mathrm{mK}$ using a tunnel diode oscillator technique operating at $14.7\,\mathrm{MHz}$ \cite{degrift1975,giannetta2023}. Each sample is mounted on a high-purity $800\,\mu\mathrm{m}$ diameter sapphire rod and fixed within the resonant coil, oriented such that $H$$\parallel$$c$ and screening currents flow along the $a$ and $b$ axes. The probe field is $<0.1\,\mu\mathrm{T}$, ensuring that the sample remains in the Meissner state. Temperature is monitored with a RuO$_2$ thermometer attached to a copper holder in which the sapphire rod is mounted. In the case that the normal state skin depth is much larger than the sample dimensions, which we expect here due to the high resistivity, measured frequency changes, $\Delta f$, are related to the effective susceptibility of the sample via $\chi_m(T) = \Delta f(T)/\alpha V_s$, where $V_s$ is the sample volume and $\alpha = 230\,\mathrm{kHz/mm^3}$ is the sensitivity of our measurement circuit. Demagnetization effects enhance the susceptibility relative to the intrinsic value, $\chi(T)$, which can be determined using $\chi(T) = \chi_m(T)/(1-\chi_m(T)N)$. In the Supplemental Material \cite{SI} we show finite element calculations of $\chi(\lambda)$ for SC cylinders with the relevant aspect ratio, and use these to determine $N$ and convert to $\Delta\lambda(T)$. A small paramagnetic background from the sapphire rod has been subtracted from the data, accounting for $<0.1\,\mathrm{Hz}$ ($\Delta\lambda\lesssim0.5\,\mathrm{nm}$). Radio-frequency self-heating was checked by varying the sample position, and therefore the magnitude of the RF field, within the coil, and found to be negligible down to the lowest temperatures measured, consistent with a low surface resistance as expected for superconductors.

Specific heat $C(T)$ was measured down to 400 mK on two samples (\#3 and \#4) using a custom-built long-relaxation calorimeter mounted on a He-3 system \cite{taylor2007}. The calorimeter utilizes a bare chip \textit{Cernox}-1030 thermometer, acting as both temperature sensor and heater. The chip is suspended in vacuum via silver-coated quartz fibres acting as electrical and thermal links. The sample is mounted on the \textit{Cernox} chip using \textit{Apiezon N} grease. The addenda of the chip, fibres and grease is determined in a measurement run prior to each sample run.

\begin{figure}
	\centering
	\includegraphics[width=0.45\linewidth]{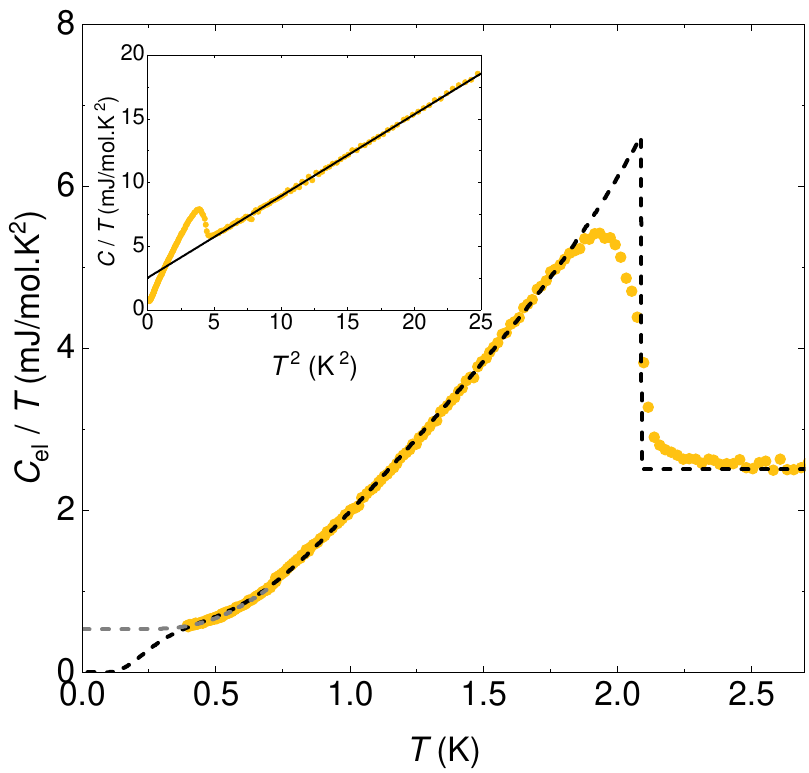}
	\caption{Main panel: The electronic specific heat of LMO (sample \#4) in the vicinity of the SC transition. The black dashed line is a fit to a specific heat model consisting of two isotropic gaps as discussed in the main text. The anomaly at $T_c$, $\Delta C/\gamma T_c \simeq 1.64$, is greater than the BCS weak-coupling value of $\Delta C/\gamma T_c = 1.43$, indicating that superconductivity in LMO is moderately coupled. The grey dashed line is an exponential fit of the data up to $T_c/3$, with a residual $\gamma_{\rm res}$ = 0.53 mJ/mol.K$^2$. Inset: Low-$T$ specific heat plotted as $C/T$ versus $T^2$. The solid line is a fit to $C/T=\gamma+ \beta T^2$.} 
	\label{fig:specificheat}
\end{figure}

\textit{Results} - The inset of Fig.~\ref{fig:specificheat} shows the specific heat $C/T$ of sample \#4 (mass = $1.54\,\mathrm{mg}$) demonstrating a bulk SC transition with $T_c\simeq2.1\,\mathrm{K}$. In contrast to the divergent resistivity, the normal state specific heat is metallic in the same temperature range, as shown by the fits to $C/T=\gamma+ \beta T^2$ above $T_c$ up to $5\,\mathrm{K}$. The parameters of the fit are $\gamma=2.50\pm0.08\,\mathrm{mJ/mol.K^2}$ and $\beta=0.64\pm0.02\,\mathrm{mJ/mol.K^4}$. The value of $\gamma\simeq2.5\,\mathrm{mJ/mol.K^2}$ determined here is more than a factor of 2 smaller than that predicted by density functional theory (DFT) calculations \cite{popovic2006}. Notably, within our experimental uncertainty, any loss of states does not appear to be $T$-dependent below $5\,\mathrm{K}$, indicating that the low-$T$ increase in the Hall coefficient and $\rho(T)$ is not due to a $T$-dependent loss of carriers \cite{chudzinski2023}.

To isolate the electronic component of the specific heat, $C_{el}/T$, we subtract the fitted phonon term, $\beta T^3$, as shown in Fig.~\ref{fig:specificheat}. At the lowest temperature measured the heat capacity has not decreased to zero in the SC state. This may reflect a strongly anisotropic or multi-gap scenario, whereby quasiparticle excitations from a smaller gap are only fully suppressed below $0.4\,\mathrm{K}$, but could also arise from low temperature nuclear contributions, or from a residual electronic term of order $\sim 20\%$ of the normal state value, associated with non-SC regions of the sample.

In the absence of data below $0.4\,\mathrm{K}$ ($\sim 0.2T_c$), and with $C_{el}/T$ remaining finite at this temperature, we are unable to resolve small gaps $\Delta\lesssim k_{\mathrm{B}}T_c$ from the specific heat data alone. Nevertheless, the magnitude of the specific heat jump, $\Delta C/\gamma T_c$, at $T_c$ provides a constraint on the overall SC energy gap scale. To proceed, we first model $C_{el}/T$ using two isotropic gaps, $\Delta_1$ and $\Delta_2$ on two 1D Fermi surface sheets, assuming weak-coupling BCS temperature dependences and calculating the entropy, $S$, using standard expressions \cite{tinkham}. The result of this fitting procedure is shown by the black dashed line in Fig.~\ref{fig:specificheat}. A representative fit is obtained for $\Delta_{\mathrm{1}} = 2.1\, k_{\mathrm{B}}T_c$, $\Delta_{\mathrm{2}} = 0.5 \, k_{\mathrm{B}}T_c$, with the relative contribution from the larger gap, $x=0.78$. However, because $C_{el}/T$ remains finite at low temperature, the value of the smaller gap should be viewed with caution. To illustrate the ambiguity, we note that the low temperature data below $T_c/3$ can also be described comparably well by a single exponential form with an additional temperature independent residual term, $C_{el}/T = \gamma_{\mathrm{res}} + A\exp(-\Delta/k_{\mathrm{B}}T)$, which yields an effective gap scale, $\Delta =2.31 \,k_{\mathrm{B}}T_c$ and $\gamma_{\mathrm{res}}=0.53\,\mathrm{mJ/mol.K^{2}}$, as shown by the grey dashed line in Fig.~\ref{fig:specificheat}.

\begin{figure}
	\centering
	\includegraphics[width=0.45\linewidth]{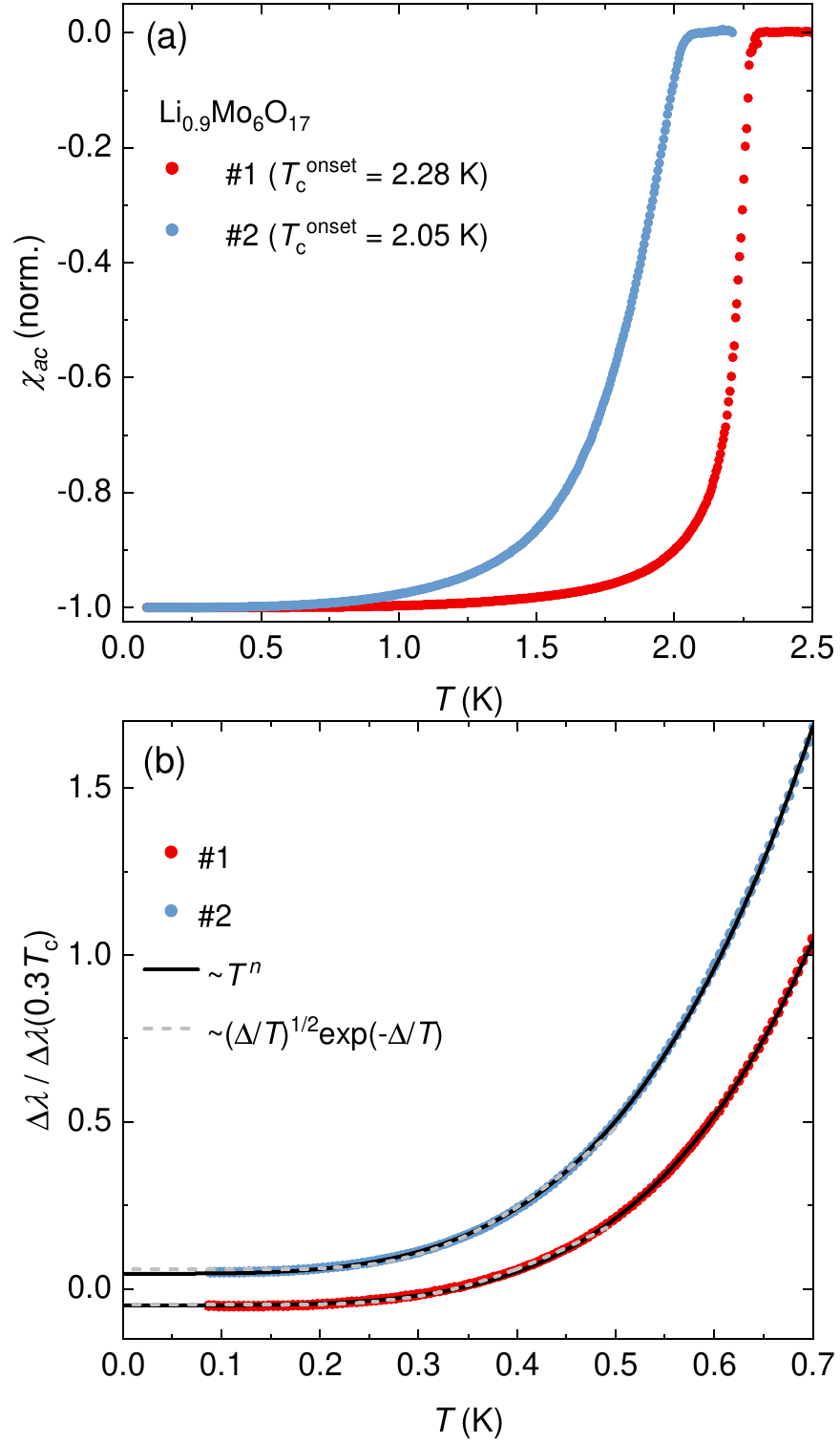}
	\caption{(a) Normalized ac susceptibility curves for \#1 and \#2 (corrected for demagnetization effects). (b) $\Delta\lambda(T)$ for $T<$ 0.3 $T_c$. Solid black curves (dashed gray lines) are fits to power-law behavior up to 0.7$\,\mathrm{K}$ (activated exponential behavior up to 0.5$\,\mathrm{K}$). The activated exponential fits return gap values of $\Delta$ = 0.82 and 0.90 $k_{\mathrm{B}} T_c$ for samples \#1 and \#2, respectively. The curves have been offset for clarity.} 
	\label{fig:pen_depth}
\end{figure}

To more accurately determine the structure of the SC energy gap, we turn to the $\Delta\lambda(T)$ measurements. The normalized ac-susceptibilities ($\chi_{\mathrm{ac}}(T)/\lvert\chi_{\mathrm{ac}}(0)\rvert$) of samples \#1 and \#2 up to $T_c$ are shown in Fig.~\ref{fig:pen_depth}(a). Both samples show a full screening fraction within the uncertainty on the demagnetisation factor. The SC transition in each case onsets at $T_c$ = 2.28 K (\#1), and 2.05 K (\#2). In the Supplemental Material we show data for sample \#3 with $T_c$ = 2.28 K \cite{SI}. The $T_c$ values for \#1 and \#3 are consistent with the highest $T_c$ reported for LMO.

\begin{figure}
	\centering
	\includegraphics[width=0.45\linewidth]{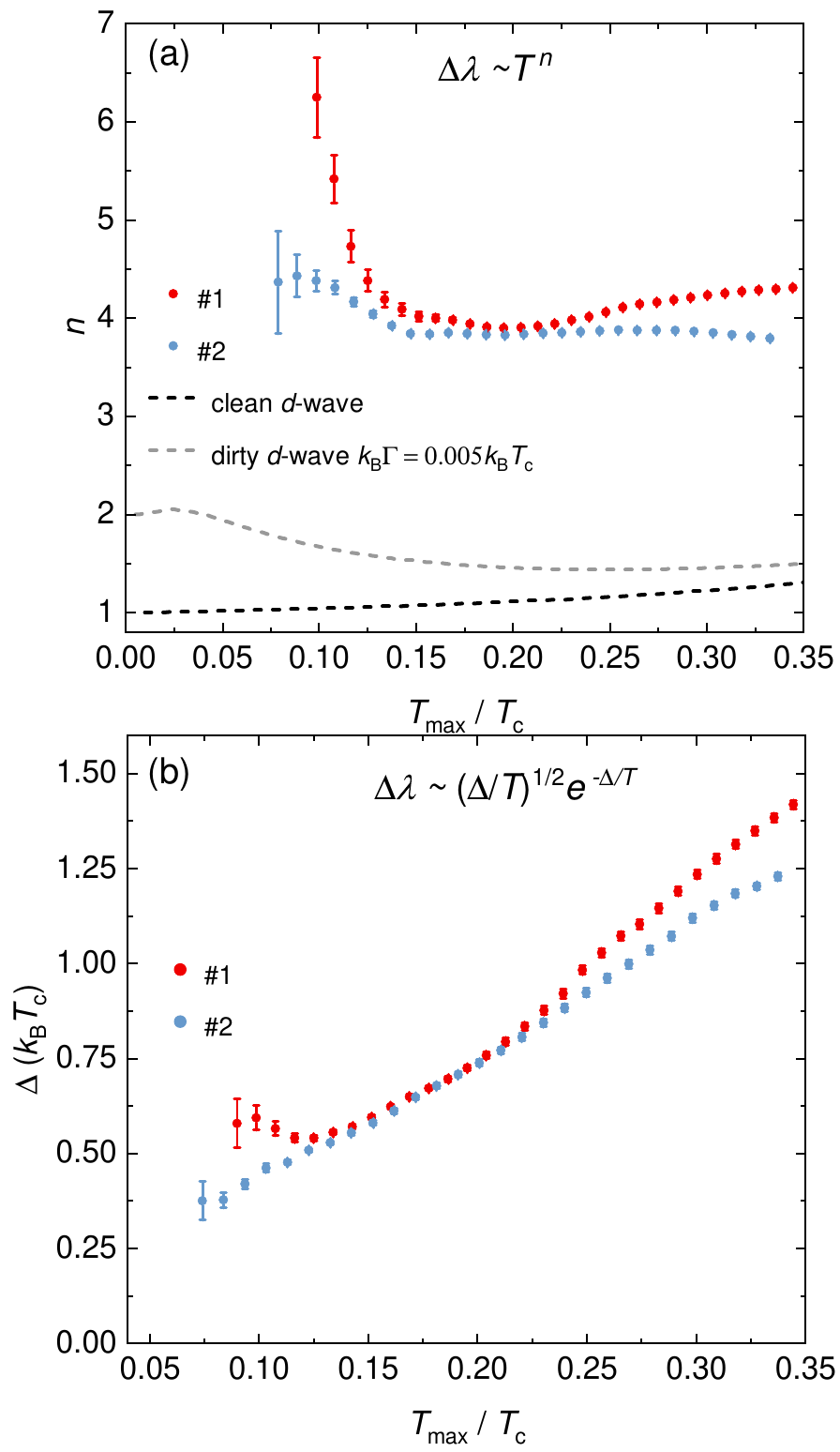}
	\caption{(a) Exponent $n$ derived from power-law fitting of the data as the upper limit of the fit, $T_{\mathrm{max}}$ is decreased. The result for a model line-nodal $d$-wave state in the clean and dirty limit ($\hbar\Gamma=0.005\,k_{\mathrm{B}}T_c$) is shown for reference. (b) Effect of decreasing $T_{\mathrm{max}}$ on $\Delta$ obtained from activated exponential fitting of the data.} 
	\label{fig:lmo_pen_depth_analysis}
\end{figure}

A comparison of the low-$T$ variation in $\Delta\lambda_{ab}(T)$ for each sample is shown in Fig.~\ref{fig:pen_depth}(b). Below $T\sim 0.3\,\mathrm{K}$, both crystals exhibit a very weak $T$-dependence consistent with a fully-gapped superconducting state. A power-law fit, $\Delta\lambda = A_1+A_2T^n$, up to $0.7\,\mathrm{K}$ yields exponents, $n = 4.25$ (\#1) and $n= 3.79$ (\#2) that are substantively larger than those expected for a nodal state in the clean or dirty limit. To check for convergent behavior as $T\rightarrow0$ we reduce progressively the upper limit of the fit, $T_{\mathrm{max}}$ \cite{grant24,cho16,kim15}. A divergence of $n$ with decreasing $T_{\mathrm{max}}$, as shown in Fig.~\ref{fig:lmo_pen_depth_analysis}(a) for both samples, strongly indicates that the response tends to being independent of $T$ at low temperature. A resultant $n\geq4$ is indistinguishable from activated exponential behavior within the experimental noise, and hence we conclude that the data are consistent with the presence of a fully-gapped state.

To estimate the magnitude of the gap, we proceed to fit the data with the activated exponential form, $\Delta\lambda = B_1+B_2(\Delta/T)^{1/2}\exp(-\Delta/T)$. In the presence of multiple or anisotropic gaps, $\Delta$ obtained from such a fit is an effective gap which converges to the minimum gap as the upper limit of the fit $T_{\mathrm{max}}\rightarrow 0$. We show example fits up to $0.5\,\mathrm{K}$ in Fig.~\ref{fig:pen_depth}(b), and show the deduced values of $\Delta$ as $T_{\mathrm{max}}$ decreases from $T_{\mathrm{max}}/T_c$ = 0.35 in Fig.~\ref{fig:lmo_pen_depth_analysis}(b). Both samples show a similar trend, in that $\Delta$ exhibits a rapid decrease with decreasing $T_{\mathrm{max}}$, and although no clear convergence occurs, the fitting procedure suggests a gap minimum $\Delta_{\mathrm{min}}\lesssim 0.5\,k_{\mathrm{B}}T_c$ for both samples. This value is much smaller than the weak-coupling value of $\Delta=1.76\,k_{\mathrm{B}}T_c$, suggesting that the gap must have substantial variations in either real or reciprocal space. Results of this analysis procedure -- performed on both nodal and nodeless models of the superfluid density -- are shown in the Supplemental Material \cite{SI}. 

In order to gain further insight into the structure of the gap, we proceed to model the superfluid density $\rho_s(T) = \lambda^2(0)/\lambda^2(T)$. This method takes into account the possible presence of multiple or anisotropic gaps, and their temperature dependence. The analysis requires knowledge of $\lambda(0)$ which is yet to be experimentally determined. Due to the large electronic anisotropy in LMO \cite{mercure2012}, we expect the measured response to be dominated by $\lambda_a(0)$. In the Supplemental Material, we present a tight-binding model parameterization of the electronic structure based on the calculations of Refs.~\cite{merino2012,nuss2014} and obtain $\lambda_b(0)\simeq0.24\,\mu\mathrm{m}$ and $\lambda_a(0)\simeq3.0 -6.5\,\mu\mathrm{m}$, depending on the tight-binding parameters used \cite{SI}. The calculated value of the Sommerfeld coefficient based on this tight-binding parameterization, $\gamma=6\,\mathrm{mJ/mol.K^2}$, is consistent with DFT calculations \cite{popovic2006}. As mentioned above, however, this is more than double the actual experimental value. This discrepancy implies a loss of states due to gapping that would in turn lead to a subsequent reduction in $\lambda^{-2}(0)$ by the same factor. Taking this into account, we obtain a revised estimate of $\lambda_a(0) = 4.6-10.1\,\mu\mathrm{m}$. As shown in the Supplemental Material an alternative estimate from the measured susceptibility gives $\lambda_a(0)\simeq20\,\mu\mathrm{m}$ for both samples, which may suggest either additional gapping or a reduced warping of the Fermi surface \cite{SI}. For the remainder of this analysis we assume $\lambda(0) \sim \lambda_a(0) =15\,\mu\mathrm{m}$, which yields a maximum gap scale consistent with the heat capacity data. As shown in the Supplemental Material, reasonable variations in $\lambda(0)$ do not alter the low-temperature form of the superfluid density and our conclusion of a small finite gap minimum.

As with the specific heat analysis, we first model the data using two isotropic gaps on the two 1D Fermi surface sheets, with the total superfluid density, $\rho_s = x\rho_1 +(1-x)\rho_2$, being a weighted contribution of the superfluid density from each sheet. Based on the analysis presented in Fig.~\ref{fig:lmo_pen_depth_analysis}, we also consider here an anisotropic gap, where the specific form of the anisotropy can be expressed as a $k_x$-dependent gap on a single quasi-1D Fermi surface,
\begin{equation}
\label{eq:gap}
\Delta(q)=\Delta_{\mathrm{max}}\!\left[1+\delta\left(\cos\!\left(\pi(1-|q|)^{\eta}\right)-1\right)\right]\ ,
\end{equation}
where $q=k_x a/\pi \in [-1,\, 1]$. Here, the gap minimum $\Delta_{\mathrm{min}}=\Delta_{\mathrm{max}}(1-2\delta)$, and the shape of the gap about $\Delta_{\mathrm{min}}$ can be tuned via the parameter $\eta$.
The normalized superfluid density is then given by
\begin{equation}
\rho_s(T) = 1- \frac{1}{\pi t}\int^{\pi/a}_{0} \int_0^{\infty}\mathrm{sech}^2\left(\frac{\sqrt{\epsilon^2+\Delta^2(t,k_x )}}{2t}\right) dk_x d\epsilon \ ,
\end{equation} 
where $t = T/T_c$.

We apply a $1/t^2$ weighting to the data to constrain the fit more by the low-$T$ behaviour, and to avoid complications arising from the finite transition width we restrict the fit to $\sim0.9\,T_c$ and fix $T_c$ as the midpoint of the diamagnetic transitions, $2.22\,\mathrm{K}$ (\#1) and $1.83\,\mathrm{K}$ (\#2).

\begin{figure}
	\centering
	\includegraphics[width=0.45\linewidth]{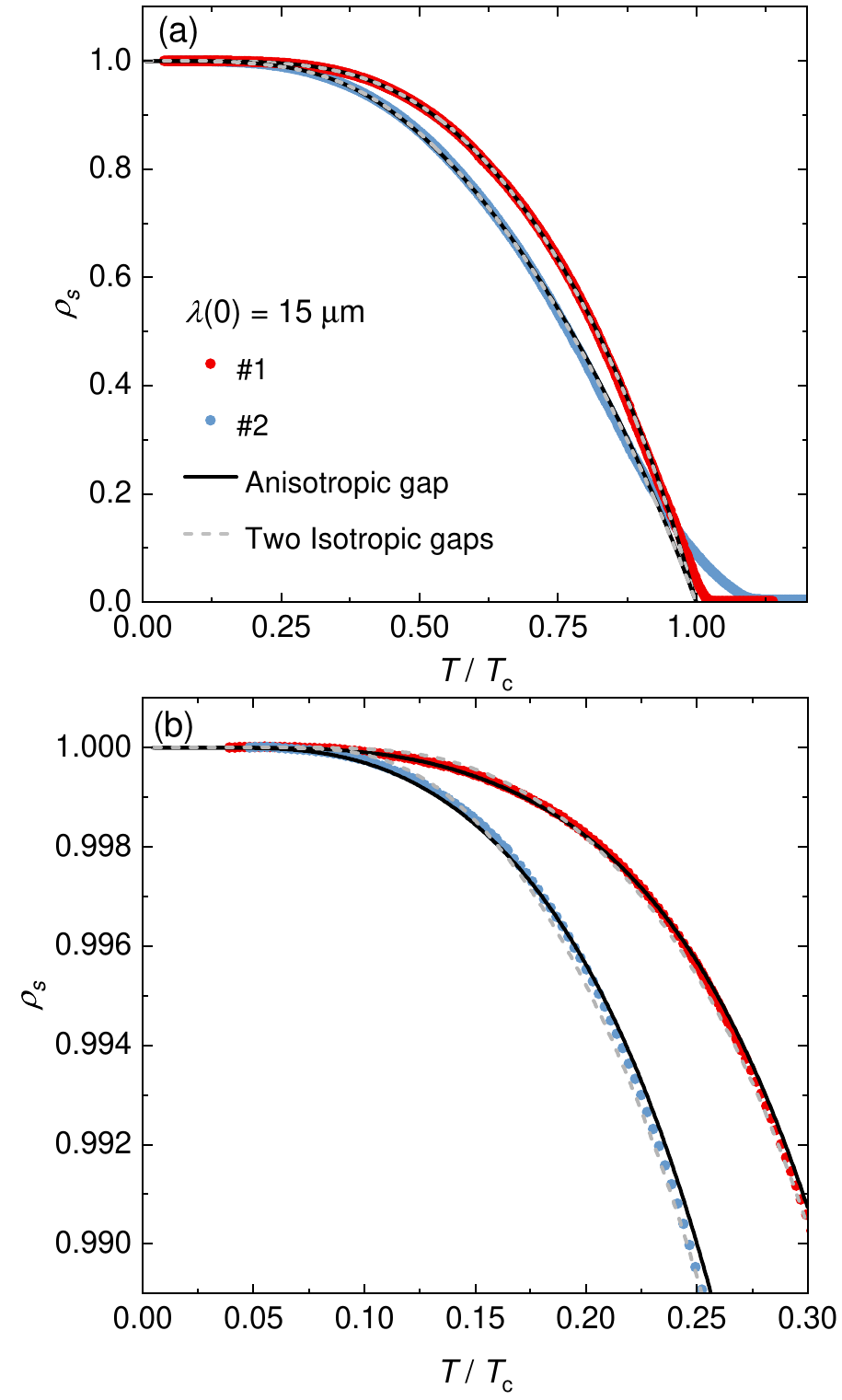}
	\caption{(a) Superfluid density of samples \#1 and \#2, determined using $\lambda_a(0)=15\,\mu\mathrm{m}$. The solid lines are fits to an anisotropic gap model as described in the main text. The dashed gray line is a fit to a two-isotropic-gap model. (b) Low-$T$ data.} 
	\label{fig:sfd}
\end{figure}

\begin{table}
	\small
	\centering
	\begin{tabularx}{0.45\textwidth}{X|X|X}
		\hline\hline
        Sample	& \textbf{\#1}  & \textbf{\#2}  \\ \hline
		\multicolumn{3}{c}{\textit{Two isotropic gaps}} \\ \hline
		{$\Delta_{1}/k_{\mathrm{B}}T_c$} & {$2.34\pm0.01$}  & {$2.03\pm0.01$}  \\
		{$\Delta_{2}/k_{\mathrm{B}}T_c$} & {$0.91\pm0.01$}  & {$0.76\pm0.01$}  \\ 
          {x} & {$0.971\pm0.001$}  & {$0.962\pm0.001$}  \\ \hline
		\multicolumn{3}{c}{\textit{Anisotropic gap}} \\ \hline
		{$\Delta_{\mathrm{max}}/k_{\mathrm{B}}T_c$} & {$2.47\pm0.01$}  & {$2.18\pm0.02$}  \\ 
		{$\delta$} & {$0.425\pm0.004$}  & {$0.431\pm0.007$} \\
		{$\eta$} & {$3.8\pm0.1$}  & {$3.3\pm0.1$} \\ 
        \hline\hline
	\end{tabularx}
    \caption{Parameters of the two-isotropic gap model and anisotropic gap model used to fit the $\rho_s(T)$ data of \#1 and \#2 using $\lambda(0) = 15\,\mu\mathrm{m}$.}\label{tb:sfd}
\end{table}

\begin{figure}
	\centering
	\includegraphics[width=0.45\linewidth]{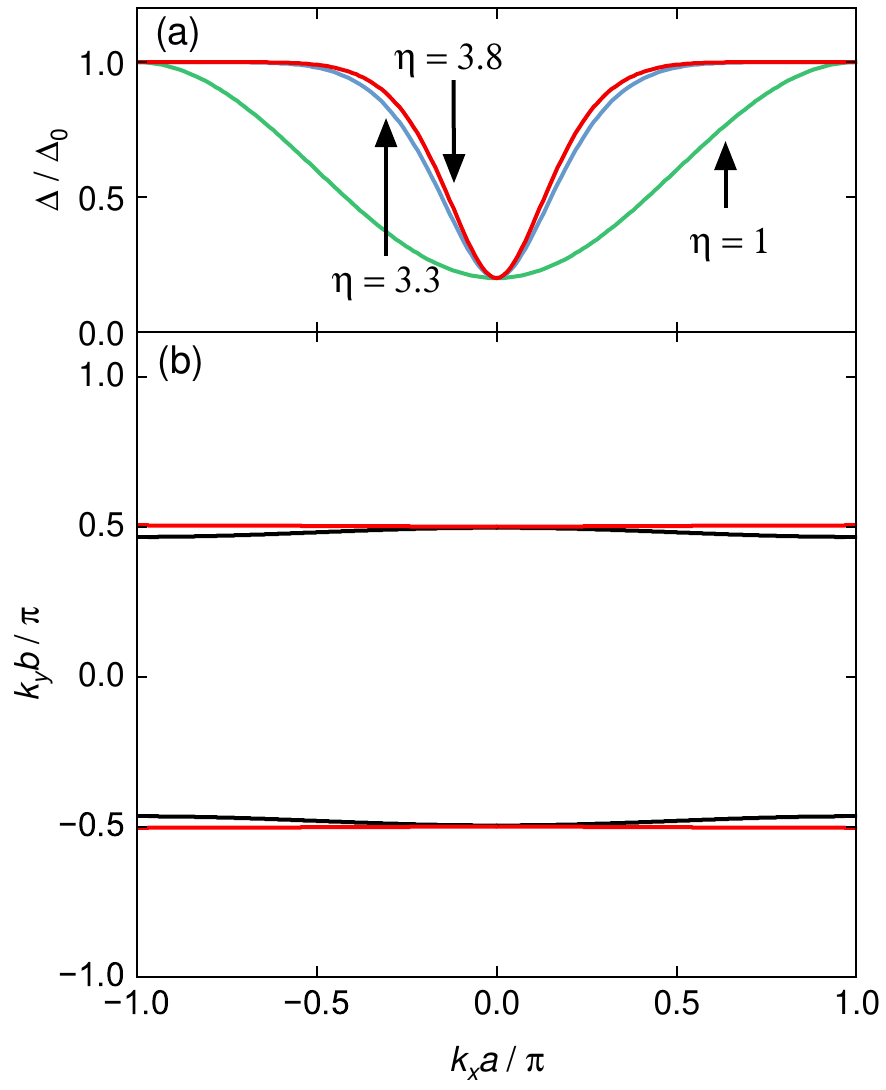}
	\caption{(a) Schematic of the anisotropic gap structure derived for $\beta=0.4$ and $\eta$ = 3.8 (\#1), 3.3 (\#2) and 1 (for comparison). Note that the locus of the gap minimum cannot be determined from this study. (b) Fermi surface of LMO as determined by the tight-binding parametrization from Ref. \cite{nuss2014}.} 
	\label{fig5}
\end{figure}

The superfluid density of \#1 and \#2 are shown in Fig.~\ref{fig:sfd}. The resultant fits (solid and dashed lines in Fig.~\ref{fig:sfd}) are in very good agreement with the data across the entire fitted temperature range. The corresponding fit parameters are shown in Table~\ref{tb:sfd}. At the lowest temperatures the two-isotropic-gap fit does not entirely capture the low temperature response. This is reflected in the fitted values of the smaller gap, $\Delta_2 /k_{\mathrm{B}}T_c = 0.91$ (\#1) and $0.76$ (\#2) which are larger than those determined from activated exponential fitting of the data in Fig.~\ref{fig:lmo_pen_depth_analysis}. For the anisotropic gap, the gap maxima, $\Delta_{\mathrm{max}}/k_{\mathrm{B}}T_c=2.47$ and $2.18$, are consistent with the larger gap determined from the specific heat and confirm that the superconductivity in LMO is moderately coupled. The gap minima are $\Delta_{\mathrm{min}}/k_{\mathrm{B}}T_c$ = 0.37 and 0.30 for \#1 and \#2, respectively. Note that these values are in good agreement with the estimates deduced from activated exponential fitting (Fig.~\ref{fig:lmo_pen_depth_analysis}) and suggests that the gap anisotropy in LMO is extremely large ($\sim 7\times$) and similar in magnitude to that reported in FeSe and CsV$_3$Sb$_5$ \cite{li2016, jiao2017, grant24}. Furthermore the anisotropy exponents $\eta =3.3$ and $3.8$ suggest narrow gap minima that occur over a very confined region in $k$-space, as shown schematically in the inset of Fig.~\ref{fig5}(a).

Both models support the notion of there being a region in momentum space that hosts a very small gap. For the two-gap model, the contribution to the total superfluid density from the smaller gap is $\sim3-4\%$, whereas in the anisotropic-gap model, the fitted $\eta$ parameter implies that the gap minima are confined to a narrow portion of the Fermi surface (see Fig.~\ref{fig5}(a)). Within the two-band interpretation, the $\sim3-4\%$ contribution to the superfluid density from the smaller gap implies that the corresponding band possesses only a very small fraction of the relevant electronic spectral weight. This picture, however, is difficult to reconcile with the electronic structure of Li$_{0.9}$Mo$_6$O$_{17}$, which, near the Fermi energy, is comprised of two nearly degenerate $d_{xy}$ bands that form two quasi-1D Fermi surface sheets (shown in Fig.~\ref{fig5}(b)), and are expected to make comparable contributions to the SC condensate. Additionally, in this case we would expect even a modest amount of interband scattering to homogenise the gap between the two sheets. By contrast, a single anisotropic gap can account for the very small inferred gap scale through narrow minima occupying only a small fraction of Fermi surface and hence is presented here as the more robust and physically plausible scenario. Again, however, the strength of the anisotropy in $\Delta$ requires that intra-band scattering is also weak. 

In order to confirm that our results are not susceptible to sample selectivity, we show in the Supplemental Material measurements of the specific heat and magnetic penetration depth obtained on a single sample (\#3). Although qualitatively in agreement with those reported above, the broader SC transition and the data quality are such that we are not able to make precise quantitative statements about the gap structure in sample \#3.

\textit{Discussion} - Our observation of activated exponential low-$T$ behavior of $\Delta\lambda(T)$ and subsequent modelling of the superfluid density reveals that the SC gap in LMO is fully-gapped yet possesses strong anisotropy. Within the $D_{2h}$ point group, our results restrict the possible pairing symmetries to the even-parity $A_{1g}$ and odd-parity $A_{1u}$, $B_{1u}$ and $B_{3u}$ states. The size and form of the gap anisotropy, shown in Fig.~\ref{fig:sfd}(b), are consistent with the $A_{1u}$ state deduced by Platt \textit{et al.}~for LMO through symmetry arguments, which is also nodeless provided that spin-orbit and interchain coupling remain small \cite{platt2016}. This specific order parameter has a symmetry-imposed sign change across $k_y$$\rightarrow$$-k_y$, coinciding with the nesting of the quasi-1D Fermi surfaces. The strong anisotropy is due to the nesting condition also being satisfied between points on a single Fermi surface sheet as well as between the inner and outer Fermi surface sheets. This results in a frustration in the sign-change and a suppression of the gap near $k_x=0$. Further investigation, for example via the introduction of disorder may be a way to confirm the sign-changing nature of the order parameter, where we expect an evolution from activated exponential to quadratic behavior, due to the pair-breaking nature of inter-sheet scattering.

Alternatively, we also consider the possibility that the apparent $k$-dependence of the SC gap reflects spatial inhomogeneity rather than intrinsic momentum-space anisotropy. In particular, the reported variability in $T_c$ across samples may indicate an underlying sensitivity to stoichiometry, disorder, or other extrinsic factors, leading to a distribution of gap values within a given sample. In such a scenario, superconductivity would be locally suppressed in certain regions of the sample, giving rise to low-energy quasiparticle excitations in the superfluid response. In this case, the presence of a markedly smaller gap necessarily implies a correspondingly reduced local $T_c$. This would be expected to produce a kink or change in curvature in the $T$-dependence of the superfluid density at the lower $T_c$, which is not observed experimentally.

Before closing, we remark that all of the above models adopt a weak-coupling, Fermi-liquid approach, with the expectation that the high-$T$ TLL state must eventually cross over to a Fermi-liquid ground state as $T$ is lowered. Presently, though, there is little evidence to support this. Photoemission studies show no evidence of a Fermi edge down to the lowest temperatures measured ($\sim 4\,\mathrm{K}$), while the power-law density of states observed by scanning tunnelling spectroscopy extends at least down to $5\,\mathrm{K}$ \cite{podlich2012}. Moreover, a recent magnetotransport study concluded that LMO follows a TLL description down to $T_c$ \cite{chudzinski2023}. Hence, the possibility that superconductivity in LMO emerges from a TLL state cannot be overlooked. Finally, given the (indirect) evidence for exciton formation in the normal state, one should also consider the interplay of the excitonic order with superconductivity. The observed anti-correlation between $T_c$ and the magnitude of the resistive upturn \cite{matsuda1986} suggests that the excitonic order is likely to have a detrimental effect on superconductivity, and since the exciton formation is localized in $k$-space to midway along the $P-K$ symmetry line \cite{lu2019}, one can speculate that the gap minima may also occur around this point.

In summary, low-$T$ $\Delta\lambda(T)$ measurements in LMO are found to be consistent with an anisotropic fully-gapped SC state. The gap minima obtained from modelling of the superfluid density are $\Delta_{\mathrm{min}}\simeq0.4\, k_{\mathrm{B}}T_c$. Supporting measurements of the specific heat show an anomaly larger than the BCS weak-coupling value, indicative of moderately-coupled superconductivity. Our results are consistent with calculations of electronic instabilities in this system, which find a highly anisotropic spin-triplet order parameter, though further experimental investigations are needed to identify the precise spin-state.

\end{document}